\documentclass[lettersize,journal]{IEEEtran}
\usepackage{amsmath,amsfonts}
\usepackage{algorithmic}
\usepackage{algorithm}
\usepackage{array}
\usepackage[caption=false,font=normalsize,labelfont=sf,textfont=sf]{subfig}
\usepackage{textcomp}
\usepackage{stfloats}
\usepackage{url}
\usepackage{verbatim}
\usepackage{graphicx}
\usepackage{cite}
\usepackage{subfig}
\usepackage{booktabs}
\begin{document}

\title{Contract-Inspired Contest Theory for Controllable Image Generation in Mobile Edge Metaverse}

\author{Guangyuan~Liu, Hongyang~Du$^{*}$, Jiacheng~Wang, Dusit~Niyato,~\IEEEmembership{Fellow,~IEEE},\\ and Dong~In~Kim,~\IEEEmembership{Fellow,~IEEE}%
\thanks{G.~Liu is with the College of Computing and Data Science, the Energy Research Institute @ NTU, Interdisciplinary Graduate Program, Nanyang Technological University, Singapore (e-mail: liug0022@e.ntu.edu.sg).}%
 \thanks{H.~Du is with the Department of Electrical and Electronic Engineering, University of Hong Kong, Hong Kong (e-mail: duhy@eee.hku.hk).}
\thanks{J. Wang, and D. Niyato are with the College of Computing and Data Science, Nanyang Technological University, Singapore (e-mail: jiacheng.wang@ntu.edu.sg, dniyato@ntu.edu.sg).}%
\thanks{D. I. Kim is with the Department of Electrical and Computer Engineering, Sungkyunkwan University, South Korea (e-mail: dongin@skku.edu).}
\thanks{* means the corresponding author}%
}
\maketitle

\begin{abstract}
The rapid advancement of immersive technologies has propelled the development of the Metaverse, where the convergence of virtual and physical realities necessitates the generation of high-quality, photorealistic images to enhance user experience. However, generating these images, especially through Generative Diffusion Models (GDMs), in mobile edge computing environments presents significant challenges due to the limited computing resources of edge devices and the dynamic nature of wireless networks. This paper proposes a novel framework that integrates contract-inspired contest theory, Deep Reinforcement Learning (DRL), and GDMs to optimize image generation in these resource-constrained environments. The framework addresses the critical challenges of resource allocation and semantic data transmission quality by incentivizing edge devices to efficiently transmit high-quality semantic data, which is essential for creating realistic and immersive images. The use of contest and contract theory ensures that edge devices are motivated to allocate resources effectively, while DRL dynamically adjusts to network conditions, optimizing the overall image generation process. Experimental results demonstrate that the proposed approach not only improves the quality of generated images but also achieves superior convergence speed and stability compared to traditional methods. This makes the framework particularly effective for optimizing complex resource allocation tasks in mobile edge Metaverse applications, offering enhanced performance and efficiency in creating immersive virtual environments.
\end{abstract}


\section{Introduction}
{
\IEEEPARstart{T}{he}  rapid development of immersive technologies has significantly realized the Metaverse, a collective virtual shared space created by the convergence of virtually enhanced physical reality and physically persistent virtual spaces~\cite{xu2022full,wang2023survey}. The Metaverse's foundation lies in creating visually rich and interactive environments that blur the line between physical and virtual realities~\cite{10592812}. As such, Among the various Metaverse applications, one of the most valuable is the generation of high-quality, realistic image that users can experience and immerse into the virtual environments~\cite{nyamtiga2022edge,du2023attention}.

Generative Artificial Intelligence (GenAI) models, particularly Generative Diffusion Models (GDMs), have demonstrated remarkable capabilities in producing high-quality images from semantic data inputs. GDM models systematically add and then remove noise from input data, allowing the generation of realistic and detailed images~\cite{wang2024unified}. The introduction of control mechanisms such as ControlNet has further enhanced the versatility of GDMs by allowing additional semantic inputs such as depth maps, segmentation maps, and edge detections to guide the image generation process~\cite{zhang2023adding}.

However, deploying controllable GDMs in mobile edge networks presents unique challenges. In mobile edge networks, resource allocation is a critical aspect due to the constrained computational capabilities of edge devices and the dynamic nature of wireless channels. Efficient resource allocation ensures that these limited resources are optimally utilized, balancing the trade-off between computation and communication. In the context of Metaverse image generation, this involves the transmission of semantic data from edge devices to a central server where high-quality images are generated~\cite{semcomliu}. Given the variability in wireless channel conditions and the need for high-fidelity semantic inputs, it is crucial to manage the allocation of transmit power and bandwidth effectively. This not only enhances the efficiency of the image generation process but also ensures an optimal user experience within the Metaverse. Existing researches~\cite{wang2023semantic,zheng2024energy} have explored various techniques for resource allocation in similar contexts, underscoring the importance of this challenge in edge computing and wireless networks.

Addressing these challenges requires efficient resource allocation and incentive mechanisms. Contest theory and contract theory provide robust mathematical frameworks for optimizing resource allocation and incentivizing edge devices to participate in the semantic data transmission process~\cite{lin2023unified,lin2023blockchain}. Contest theory models the competitive behavior among multiple agents competing for limited resources, while contract theory facilitates the design of agreements that align the incentives of the central server and edge devices~\cite{xu2023contract,liu2023vision}. However, these traditional incentive mechanisms often assume stable communication conditions and homogeneous agent capabilities, which are not always feasible in mobile edge networks where channel conditions and device capabilities can vary significantly. This variability can lead to inefficiencies and suboptimal outcomes if not properly managed. 

In this paper, we propose a contract-inspired contest framework that is tailored to address these specific challenges by dynamically adjusting incentives and resource allocations in response to the fluctuating network conditions and diverse device capabilities, ensuring more effective and efficient operations within the mobile edge Metaverse environment. The contract-inspired contest theory incentive mechanism introduces a complex multi-tier optimization problem for resource allocation. To efficiently solve this problem, we implement an innovative approach that combines GDMs with Deep Reinforcement Learning (DRL), leveraging the strengths of both techniques to optimize the incentive mechanism in dynamic mobile edge environments. The contributions of this paper are summarized as follows:
\begin{itemize}
\item We introduced a mobile edge immersive Metaverse image generation framework that utilizes GDMs and advanced control mechanisms to produce high-quality images. The adoption of controllable GDMs is crucial as they offer superior image quality by systematically managing noise during the generation process, resulting in more physically persistent and immersive virtual environments.
\item We introduce a contract-inspired contest theory-based incentive mechanism to effectively tackle the challenges of resource allocation and semantic data transmission quality in edge computing. This mechanism leverages contest theory, which efficiently manages information asymmetry by influencing the behavior of edge devices without requiring direct information sharing. The integration of contract theory mitigates the risk of collusion by setting a fixed total reward, ensuring that the incentive structure remains efficient.
\item We implement an innovative approach combining GDMs with DRL to address the complex multi-tier optimization problem. This hybrid method excels in resource allocation parameter generalization, particularly in scenarios where parameters follow certain distributions, leading to faster convergence and more robust performance compared to traditional optimization techniques.
\end{itemize}

The remainder of this paper is structured as follows: Section~\ref{S2} provides a comprehensive review of related works, focusing on GDMs and their application in mobile edge networks, along with incentive mechanisms in mobile edge computing. Section~\ref{S3} introduces the system model and outlines the key components of the proposed framework. Section~\ref{S4} delves into the problem formulation, presenting the contract-inspired contest theoretic multi-agent model. Section~\ref{S5} discusses the integration of DRL with GDMs to solve the proposed framework. Section~\ref{section:na} offers detailed numerical analysis and experimental results to validate the effectiveness of the approach. Finally, Section \ref{conclu} concludes the paper with a summary of findings and potential future directions.
\section{Related Work}\label{S2}

\subsection{Image Generation and Its Application in Mobile Edge Networks}

GDMs are advanced models that transform random noise into high-quality images through a denoising process. These models systematically degrade the input data by adding Gaussian noise and then learn to reverse this process to restore the original data through incremental denoising and reconstruction~\cite{wang2024unified}. This technique is employed by several state-of-the-art text-to-image generation models such as Stable Diffusion~\cite{stablediffusion}, Disco-Diffusion~\cite{discodiffusion}, and DALL-E2~\cite{marcus2022very}, which have demonstrated remarkable performance in producing high-quality photorealistic images.

ControlNet extends the capabilities of diffusion models by introducing mechanisms to control image generation using additional modalities~\cite{zhang2023adding}. For instance, ControlNet allows the incorporation of various  ``semantics" such as depth maps, segmentation maps, open pose estimations, and edge detections to guide the image generation process. This ensures that the generated images adhere closely to the desired attributes and constraints defined by these additional inputs.

Deploying diffusion models and ControlNet in mobile edge networks presents unique challenges and opportunities. Edge devices typically have limited computing resources compared to centralized servers. To address these challenges, several strategies have been proposed to enhance the feasibility of on-device deployment and model reduction. One effective approach is quantization and profiling, which optimizes the models for edge devices by reducing their computational complexity without significantly compromising performance~\cite{castells2024edgefusion}. For instance, the SnapFusion model can generate 512×512 images from text prompts on mobile devices in under two seconds, demonstrating the viability of diffusion models for immersive mobile edge Metaverse applications~\cite{li2024snapfusion}. Another common approach proposed by the authors in~\cite{du2023exploring} is to leverage the strengths of both central and edge infrastructures by distributing the diffusion process across these components. However, these scenarios may not be suitable for controlled diffusion methods. Typically, controlled diffusion methods involve both the diffusion model and the control mechanism, which require more integrated and coordinated processing.

In controlled diffusion scenarios, a prevalent strategy involves edge devices handling the initial data processing and transmission of semantic information, while central servers perform the intensive denoising computations required to generate high-quality images~\cite{semcomliu}. This method is particularly suitable for applications where both control extraction and diffusion are essential~\cite{zhang2023adding}. 

\subsection{Incentive Mechanism in Mobile Edge Computing}

In edge computing environments, multiple agents typically need to collaborate and make decisions to optimize the use of limited computing resources. Mobile edge computing leverage theories and frameworks such as contest theory and contract theory to facilitate effective coordination and resource management among these devices.

Contest theory has been used to model situations where multiple agents compete for a limited set of resources or rewards~\cite{contest,7287773}. In edge computing, contest theory can be applied to allocate computing resources among various tasks and devices. The competition is designed such that agents are incentivized to exert optimal effort to win a share of the resources, thereby enhancing overall system performance. 

By applying contest theory, edge computing systems can effectively manage resource allocation in a decentralized manner. For instance, the authors in~\cite{liu2023vision} utilized contest theory to balance the rendering power among users in the same service to improve service immersiveness. In addition, a contest-based incentive mechanism was introduced in~\cite{wang2023semantic} to maximize the efforts of participants to improve the quality of service of Metaverse service providers. While these works focus on applying contract or contest theory to improve performance, they typically assume uniform agent capabilities and static incentive structures. In contrast, our proposed contract-inspired contest framework adapts dynamically to fluctuating wireless conditions and diverse device capabilities.

On the other hand, contract theory provides a framework for designing agreements between a principal (such as a central server) and multiple agents (such as edge devices)~\cite{8402115,zhang2017multi}. In the context of edge computing, contract theory is used to create incentive-compatible contracts that ensure agents act in the best interest of the overall system.

 Contract theory can be applied to model payment plans and utilities in edge computing environments. For example, the study detailed in~\cite{du2023attention} explores how contract theory can be used to design optimal contracts between mobile service providers (MSPs) and infrastructure providers (InPs), ensuring both incentive compatibility (IC) and individual rationality (IR) constraints are met. These contracts help align the interests of MSPs and InPs, promoting efficient resource utilization and improved service quality. 
 
 Additionally, the work in~\cite{du2023ai} investigates the application of diffusion models to enhance contract design. The study proposes an AI-generated contract method using a conditional diffusion model to iteratively refine initial random contracts, optimizing the design to maximize utility functions. This approach leverages generative capabilities of the diffusion models to handle high-dimensional environments and complex decision-making problems effectively. However, the system may need to generate contracts separately in situations with multiple contracts with the multiple-payment receivers, which can be time-consuming and computationally intensive.
 
Existing applications of contract theory in mobile edge computing primarily consider static agent behaviors or homogeneous environments. Our innovation lies in the deep integration of contract and contest theories to design a more nuanced incentive mechanism, which prevents collusion while encouraging agents to exert optimal effort based on their respective capabilities. This ensures a suitable distribution of rewards and aligns with system goals more effectively compared to traditional winner-takes-all or fixed distribution strategies used in the existing literature.

\subsection{Enhanced DRL with GenAI in Network Optimization}

In addition to generating high-quality images, GDM are increasingly adopted for network optimization, offering robust and efficient solutions. The ability to handle complex data distributions is particularly beneficial in scenarios that require rapid, high-performanc e decision-making~\cite{du2024enhancing}. 

In~\cite{du2024enhancing}, the authors presented case studies across several significant intelligent network scenarios. The results showed that GDM-DRL models exhibited faster convergence and more stable performance than that of traditional DRL approaches such as Proximal Policy Optimization (PPO) and Soft Actor-Critic (SAC). The reward curves from GDM-DRL models indicated smaller fluctuations, highlighting their enhanced stability. The flexibility of GDMs in modelling diverse behaviors and iterative planning provides novel perspectives on decision-making processes~\cite{ching2006markov}. This flexibility is advantageous in dynamic network environments, where conditions frequently change. Additionally, GDMs improve sample efficiency and noise reduction, distinguishing signal from noise more effectively during environment exploration~\cite{zhang2024generative}. This improvement leads to better overall learning results, making GDMs well-suited for complex network optimization tasks.

A notable application of GDMs within DRL frameworks is in adjusting the contention window and frame length jointly to ensure efficient and reliable Wi-Fi communication~\cite{liu2024gdmdrl}. By leveraging GDMs' generative capabilities, the DRL framework can make optimal adjustment under varying conditions, demonstrating the practical utility of this integration in real-world scenarios~\cite{du2024enhancing}. 
\begin{figure*}[t!]
\centering{\includegraphics[width=0.9\textwidth]{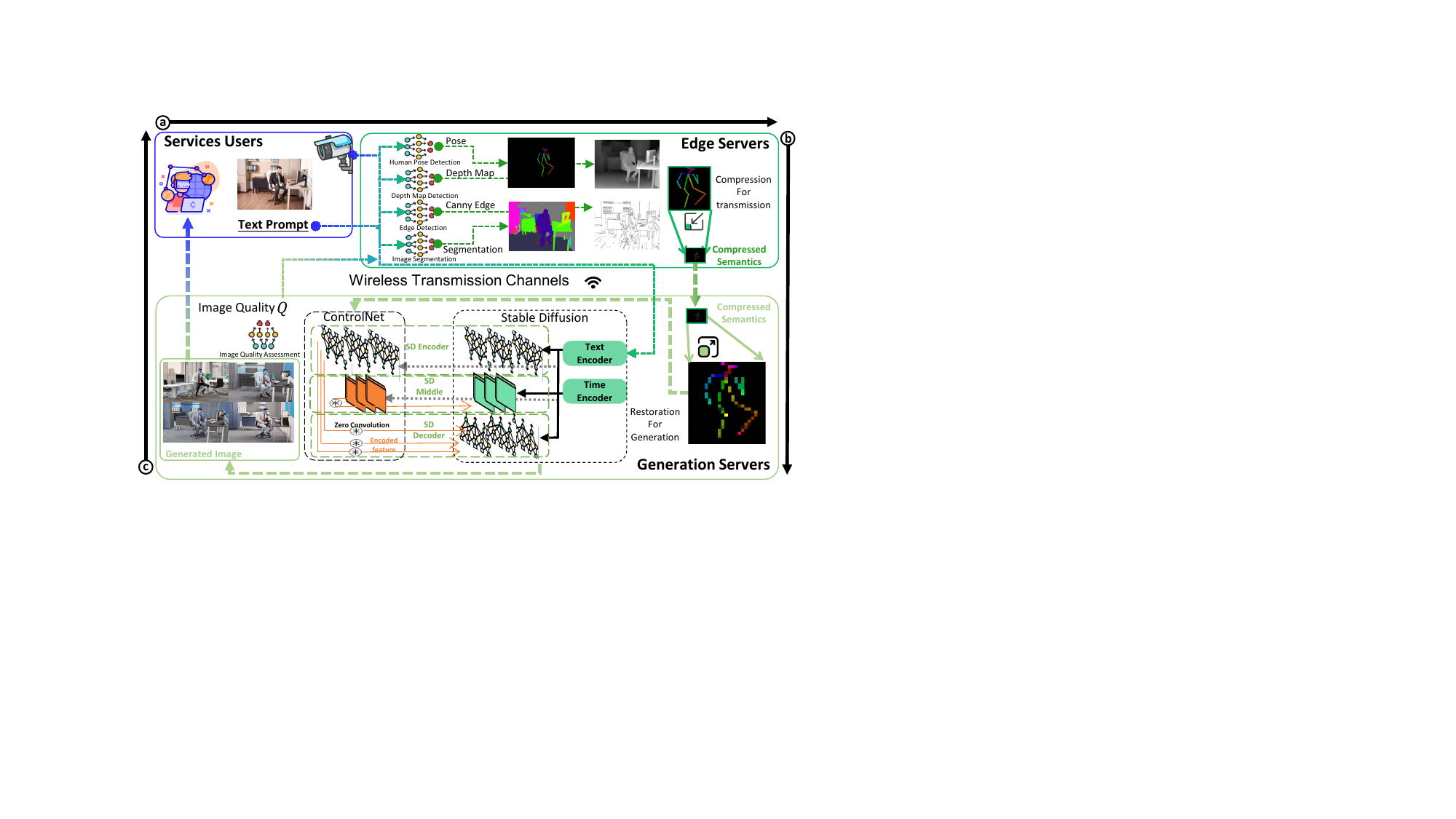}}
\caption{System components and data flow for the mobile edge immersive Metaverse image generation. The data flow in the system model is in the following steps: \textbf{(a)} User devices capture input images and upload them to an edge server for semantic extraction. \textbf{(b)} The edge server extracts semantics for different tasks and transmits the semantics to a generation server. Before transmission, semantics are compressed in different level to be transmitted at the same time. \textbf{(c)} The generation server receives and recovers the semantics, generates the images, and then sends them back to the users.}
\label{systemodel}
\end{figure*}

GDMs provide high-quality data generation and iterative refinement capabilities, while DRL frameworks ensure efficient and adaptive decision-making. However, traditional DRL methods like PPO and SAC often experience slow convergence and performance instability, especially in dynamic environments with complex data distributions. While integrating GDM with DRL provide faster convergence and better sample efficiency in various network optimization tasks, our work uniquely extends these benefits to multi-agent competitive environments in mobile edge networks, which has not been done before in the literature. By coupling GDM with contest incentives, we ensure that resource allocation decisions not only converge quickly but also adapt effectively to changing conditions in real-time. This is particularly valuable in the context of Metaverse applications, where maintaining high-quality image generation while managing dynamic competition among edge devices is crucial.

\section{System Model}\label{S3}

In this section, we present the system model for enhancing mobile edge immersive Metaverse image generation using a contract-inspired contest theoretic approach. Our proposed system is designed to efficiently allocate resources and optimize the generation of high-quality image generation in a multi-agent environment. The system consists of the key components and their interactions, as well as the assumptions and parameters used in the model.

\subsection{System Components}

As illustrated in Fig.~\ref{systemodel}, a typical mobile edge immersive Metaverse image generation system contains the following components:
\subsubsection{User Devices}           

User devices are equipped with cameras and computing resources to capture input images. These devices upload the captured images to the edge server for further processing. Each user device subscribes to the image generation service and pays a subscription fee, denoted by \( b_i \) per image transfer task, to access the service.

\subsubsection{Edge Server}

The edge server is responsible for extracting semantic information from the input images user devices provide. The extracted semantics include various features such as depth maps, segmentation maps, pose estimations, and edge detections. These semantics are essential for the image generation process and are transmitted from the edge server to the generation server, which is often located in the cloud. Given that the edge server can be mobile, and hence requires wireless transmission to the generation server in the cloud which introduces variability in bandwidth and latency, making it challenging to maintain the fidelity of the semantic data during transfer. These factors are critical in determining the overall performance of the image generation service, as both the available bandwidth and the target latency of the backhaul link between the edge server and the generation server are limited. Additionally, the amount of semantic data required for image generation is variable, depending on the complexity of the target image. Therefore, optimizing the semantic communication process is essential to balance resource consumption and ensure that high-quality semantic data can be transmitted efficiently to generate realistic and immersive images in the Metaverse environment~\cite{semcomliu,zhang2023adding}. The edge server allocates transmit power \( P_i \) for each semantic transfer task, in order to balance the quality of the transmitted semantics and resource consumption.

\subsubsection{Generation Server}

The generation server receives and utilizes the semantic data from the edge server to generate high-quality images. The quality of the generated images, denoted by \( Q_i \), depends on the quality of the semantic data received. The generation server implements a payment plan based on image quality and allocates resources accordingly to maximize overall utility.

\begin{table*}[ht]
\centering
\caption{Summary of Notations and Acronyms}
\begin{tabular}{|c|l|c|l|}
\toprule
\hline
\textbf{Notation} & \textbf{Description} & \textbf{Notation} & \textbf{Description} \\ \toprule
\multicolumn{4}{|c|}{\textbf{System Model Notations}} \\ \hline
\( a_i \)        & Capability of semantic transfer task \( i \) & \( Q_{reward,i} \) & Image reward score for image \( i \) \\ \hline
\( b_f \)        & Constant base payment to the edge server & \( Q_{SSIM,i} \) & SSIM score for image \( i \) \\ \hline
\( b_i \)        & Subscription fee for user \( i \) & \( r_i \)        & Reward for semantic transfer task \( i \) \\ \hline
\( B \)          & Bandwidth of a wireless channel & \( R_g \)        & Total reward based on image quality \\ \hline
\( C(a_i, P_i) \) & Cost function for task \( i \) based on transmit power & \( S_i \)        & Size of semantic data for task \( i \) \\ \hline
\( D \)          & Data rate of the channel & \( T \)          & Steps in the diffusion process \\ \hline
\( I_{edge} \)   & Total payment to the edge server & \( Z(D(P_i)) \)  & Compression level for task \( i \) \\ \hline
\( N_u \)        & Number of semantic transfer tasks/devices & \( \gamma_{th} \) & SNR threshold for outage \\ \hline
\( P_i \)        & Transmit power for task \( i \) & \( \theta \)      & Outage probability threshold \\ \hline
\( P_{total} \)  & Total transmit power budget & \( u_g \) & Unit fee per quality of generated image \\ \hline
\( u_e \) & Additional fee per image quality unit & \( \sigma_N^2 \) & Noise power \\ \toprule
\multicolumn{4}{|c|}{\textbf{Notations for Diffusion Process}} \\ \hline
\( \alpha_t \)   & Weighting factor in forward diffusion & \( \epsilon_0, \epsilon_\theta \) & Noise prediction function \\ \hline
\( \beta_t \)    & Variance added during forward diffusion & \( \mu_t, \Sigma_t \) & Mean and covariance in diffusion \\ \hline
\( \bar{\alpha}_t \) & Cumulative product of \(\alpha_t\) & \( q(y_t | y_0) \) & Distribution of noise-added data \\ \hline
\( y_0 \)        & Original data (payment and reward parameters) & \( y_t \)        & Noisy data at step \( t \) \\ \toprule
\multicolumn{4}{|c|}{\textbf{Notations for DRL Process}} \\ \hline
\( A_0 \)        & Optimal action parameters & \( M \)        & Total number of diffusion steps \\ \hline
\( e_t \)        & Environment state at time \( t \) & \( N_b \)        & Batch size for training \\ \hline
\( \chi, \psi \) & Weights of utility and parameter networks & \( \eta \)        & Soft target update parameter \\ \hline
\( \mathcal{B} \) & Replay buffer in DRL & \( \lambda\)     & Discount factor in DRL \\ \hline
\( Q_{\chi} \)   & Utility network with weights \( \chi \) & \( \zeta \)     & Exploration noise \\ \hline
\( \mathcal{C} \)   & Number of possible actions in DRL & \( \mathcal{S} \)     & Number of possible states in DRL \\ \hline
\multicolumn{4}{|c|}{\textbf{Acronyms}} \\ \hline
\textbf{AR} & Augmented Reality & \textbf{DRL} & Deep Reinforcement Learning \\ \hline
\textbf{GenAI} & Generative Artificial Intelligence & \textbf{GDM} & Generative Diffusion Model \\ \hline
\textbf{IC} & Incentive Compatibility & \textbf{InP} & Infrastructure Provider \\ \hline
\textbf{IR} & Individual Rationality & \textbf{MIDAS} & Multi-Instance Depth Attention Sampling \\ \hline
\textbf{MSP} & Mobile Service Provider & \textbf{PPO} & Proximal Policy Optimization \\ \hline
\textbf{SAC} & Soft Actor-Critic & \textbf{SNR} & Signal-to-Noise Ratio \\ \hline
\textbf{SSIM} & Structural Similarity Index Measure & \textbf{VR} & Virtual Reality \\ \hline
\bottomrule
\bottomrule
\end{tabular}
\label{tab:notations}
\end{table*}

\subsection{Communication Model}

The communication model between the edge server and the generation server involves the transmission of semantic data over wireless channels. The quality of the transmitted semantic data is influenced by the compression level which is limited by the achievable data rate. As the data rate $D$ limited by the Shannon-Hartley theorem under Rayleigh fading~\cite{somekh2000shannon} can be expressed as follows:
\begin{equation}\label{capacity}
    D\left(P_i\right) = B \log_2\left(1 + \frac{P_i|g|^2}{\sigma_{N}^{2}}\right),
\end{equation}
where $B$ represents the channel bandwidth, and SNR denotes the signal-to-noise ratio (SNR). In Rayleigh fading, SNR is a random variable due to the varying channel conditions caused by multipath effects. $P_i$ is the transmit power of semantic transfer task $i$, $|g|^2$ indicates the channel power gain which follows a random exponential distribution, and $\sigma_{N}^{2}$ denotes the noise power.
\begin{figure}[t!]
\centering{\includegraphics[width=0.5\textwidth]{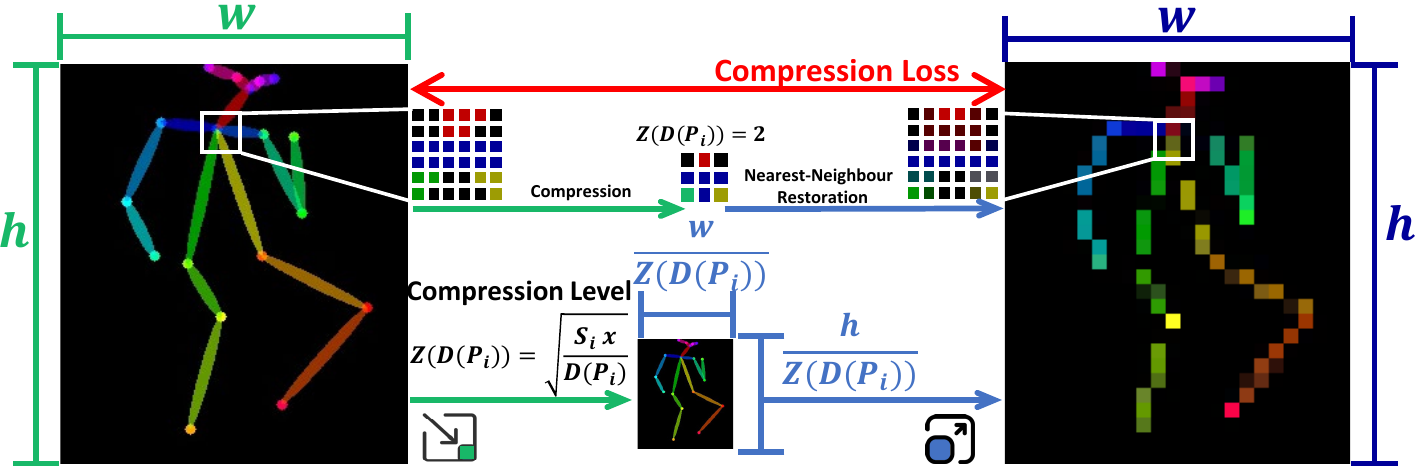}}
\caption{Illustration of the relationship between compression level and the corresponding compression loss. As the compression level $Z\left(D\left(P_i\right)\right)$ increases, the width $w$ and height $h$ of the semantic data are downscaled, leading to a decrease in the amount of transmitted information. However, this results in a loss of detail in the semantic data.}
\label{comrpessionerror}
\end{figure}
A lower data rate means that less information can be transmitted within a given time frame. Consequently, semantic data must be compressed or scaled down to fit the available data rate, leading to a loss of detailed information and degraded semantic quality as shown in Fig.~\ref{comrpessionerror}. This directly affects the quality of the generated images. On the other hand, high data rate scenarios allow for more simultaneous data transmission, reducing the need for compression or downscaling thus enables the transmission of richer, more detailed semantic data, enhancing image quality. Therefore, higher channel capacity and data rate are crucial for maintaining the integrity and quality of semantic data transmitted from edge server to generation server, ultimately leading to superior image quality in immersive Metaverse applications.

To achieve real-time performance, the compression level of the semantic transfer task must account for both the channel capacity and the reliability of transmission. In wireless communication systems, the outage probability, $P_{out}$, plays a crucial role in determining the effective data rate. The outage probability is a function of the transmit power $P_i$ and can be derived considering a Rayleigh fading channel model~\cite{xie2020lite}. The system is considered in an outage when the SNR falls below a certain threshold, denoted as $\gamma_{th}$. The probability of outage can be expressed as~\cite{moinuddin2013simple}:
\begin{equation}\label{poutpi}
P_{out}\left(P_i\right) = 1 - \exp\left(-\frac{\gamma_{th} \sigma_{N}^{2}}{P_i}\right).
\end{equation}
This formula assumes that the channel gain $|g|^2$ follows an exponential distribution, characteristic of Rayleigh fading, with a mean value of 1. The outage probability $P_{\text{out}}\left(P_i\right)$ indicates that the instantaneous SNR $\gamma$ falls below a threshold $\gamma_{\text{th}}$, where the SNR $\gamma$ is given by:
\begin{equation}
\gamma = \frac{P_i |g|^2}{\sigma_{N}^{2}}.
\end{equation}
Considering the impact of outage probability on the effective data rate, we can define the compression level to ensure that each service satisfies a minimum frame rate requirement of $x$ frames per second (fps), such as 30 fps, which is a common threshold for real-time services~\cite{tan2020real}. The compression level, accounting for potential retransmissions due to outages, can be obtained by:
\begin{equation}
Z\left(D\left(P_i\right)\right) = \sqrt{\frac{S_i \times x}{D\left(P_i\right) \times \left(1 - P_{out}\left(P_i\right)\right)}},
\end{equation}
where $S_i$ represents the size of the semantic data, $D\left(P_i\right)$ is the data rate of the channel for task $i$, and $Z\left(D\left(P_i\right)\right)$ represents the compression level (by both width and height) for task $i$. The term $\left(1 - P_{out}\left(P_i\right)\right)$ accounts for the probability of successful transmission.

\subsection{Image Quality Metrics}

The quality of the generated images is evaluated using two metrics:

\begin{itemize}
    \item \textbf{Structural Similarity Index Measure (SSIM)~\cite{brunet2011mathematical}:} This metric quantifies the semantic alignment between the generated image and the intended content, ensuring adherence to specified constraints and the immersiveness.
    \item \textbf{Image Reward~\cite{xu2024imagereward}:} This metric simulates human preferences in image quality assessment by incorporating a model that predicts human evaluative responses to images.
\end{itemize}
The generated images are assessed using the Image Reward \(Q_{reward}\) and SSIM similarity score \(Q_{SSIM}\), which are both normalized to $[0, 1]$. The scores are combined as follows:

\begin{equation}\label{eq:quality}
Q_i = \beta \cdot Q_{reward,i} + \left(1-\beta\right) \cdot Q_{SSIM,i}.
\end{equation}
The overall image quality \( Q_i \) is a weighted combination of these metrics, which is influenced by the data rate $D$ limited by the channel capacity. Increased channel data rate reduces the need for data compression or downscaling, ensuring that more detailed and high-quality semantics are transmitted~\cite{semcomliu}.

\subsection{Semantic Types}

The semantics deployed for image generation tasks include Multi-Instance Depth maps via Attention Sampling (MIDAS), segmentation, pose estimation, and Canny edge detection~\cite{semcomliu}:

\begin{itemize}
\item \textbf{Depth Map:} This semantic measures the distance from the imaging sensor to each pixel's corresponding point in the real world, rendering this distance as a grayscale image. The MIDAS algorithm enhances this by using a novel attention mechanism to generate high-quality depth maps from 2D images~\cite{fuster2022nested}. Depth maps are particularly useful in applications requiring spatial awareness, such as augmented reality (AR), robotics, and 3D reconstruction, where understanding the distance between objects is crucial for creating immersive and interactive environments.

\item \textbf{Segmentation:} This semantic divides an image into regions or objects, facilitating further analysis~\cite{minaee2021image}. Segmentation is crucial for immersive Metaverse applications, where creating realistic and interactive virtual environments requires precise identification and manipulation of different objects within a scene. For example, in virtual reality (VR) or AR, segmentation allows for the accurate overlay of virtual objects onto real-world environments, enabling seamless interactions and enhancing the user's sense of immersion. 

\item \textbf{Canny Edge Detection:} This semantic is a multi-stage algorithm that identifies a wide range of image edges, producing a binary output that highlights the boundaries of objects within the image~\cite{xie2015holistically}. In the context of immersive Metaverse applications, Canny edge detection plays a vital role in defining sharp and clear boundaries of objects, which is essential for rendering high-quality, detailed visuals. This clarity enhances the realism of the virtual environment, making objects more distinguishable and interactive. 

\item \textbf{Pose Estimation:} This semantic is particularly relevant when dealing with images of human figures, identifying body postures and extracting skeleton position data~\cite{osokin2018real}. Pose estimation is widely used in applications such as motion capture for animation and gaming, fitness tracking, and human-computer interaction, where understanding human movement and body orientation is essential.
\end{itemize}

The system model highlights the intricate interplay between user devices, edge servers, and generation servers in a mobile edge immersive Metaverse environment. Based on this system model, next we delve into the problem formulation for optimal resource allocation, leveraging contract-inspired contest theory to address the challenges inherent in mobile edge computing environments.
}
\section{Contract-Inspired Contest Theory Based Multi-Agent Problem formulation}\label{S4}

\begin{figure}[t!]
\centering{\includegraphics[width=0.48\textwidth]{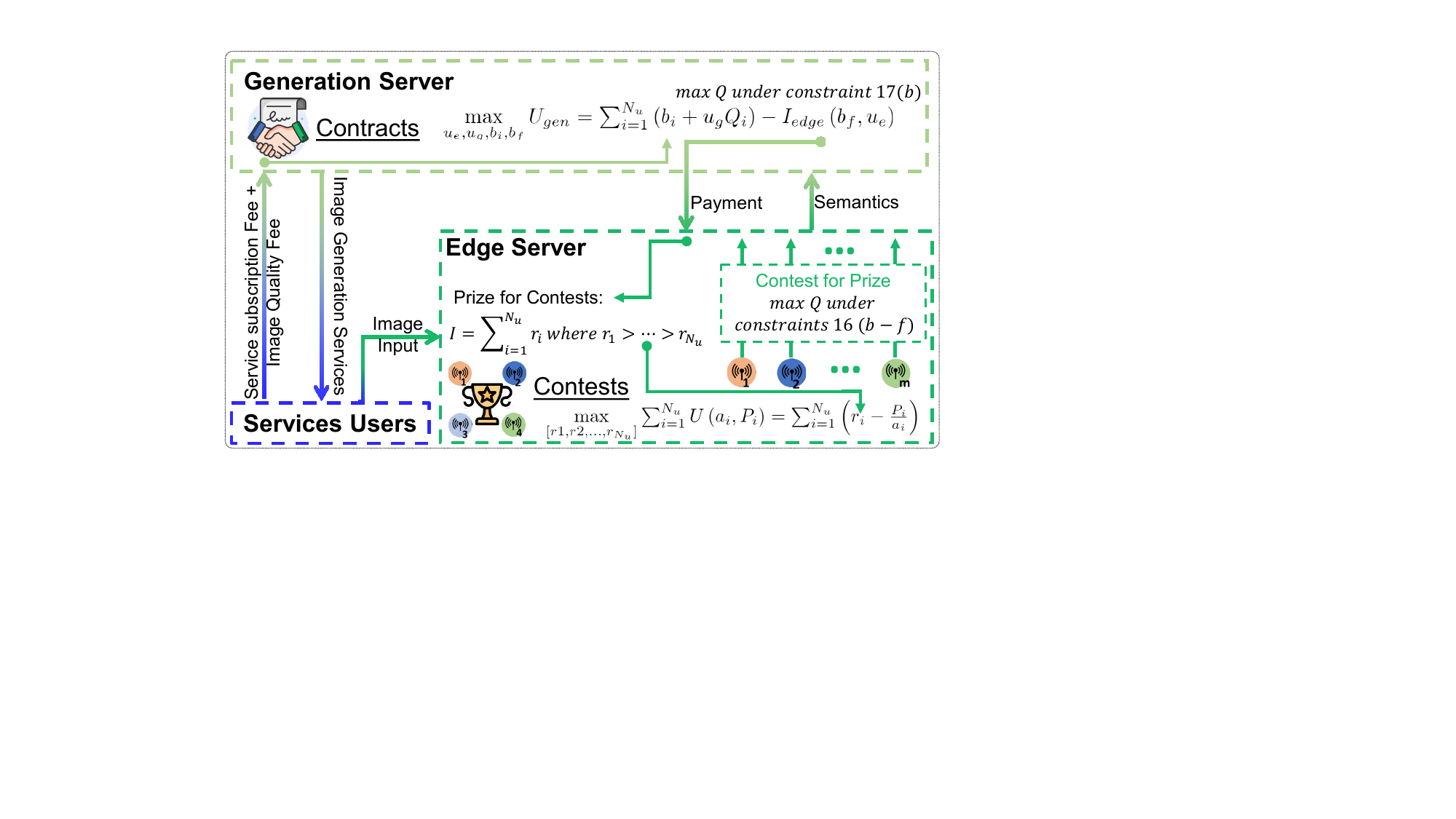}}
\caption{System model for the proposed mobile edge immersive Metaverse image generation framework, detailing the flow of information and relationships between each component. User devices capture input images, which are transmitted to the edge server for semantic extraction. The edge server extracts various types of semantic information and forwards them to the generation server, which then generates high-quality images. The edge server runs a contest-based mechanism to incentivize the semantic transfer tasks by allocating transmit power based on their contributions. Meanwhile, the generation server sets a payment plan to effectively adjust the overall performance.}
\label{systemdigram}
\end{figure}

In response to the aforementioned challenges, we propose a contract-inspired contest theory incentive mechanism as illustrated in Fig.~\ref{systemdigram}. In our proposed framework, contract-inspired contest theory defines an initial incentive mechanism, where both the award distribution scheme and total award scheme are dynamically adjusted by the DRL agent. This allows the system to efficiently incentivize participants while adapting to dynamic changes in the mobile edge network environment. In this section, we present the formulations of the contract-inspired contest incentive mechanism framework. 

\subsection{Incentive in Generation Server}
Inspired by contract theory, we propose a payment plan between the generation server and the edge server in this subsection. The utility of the generation server is formulated by considering the generated image quality.
\subsubsection{Generation Payment Plan}
To fully unlock the potential of wireless networks, a well-structured payment plan is essential. This plan should ensure that all image generation services can effectively benefit from tasks involving semantic transfer. We propose a contract theory inspired payment plan where an edge server receives a payment of award setting from the generation server according to the generated image quality which is influenced by the received semantic quality. The payment function can be expressed as follows:
\begin{equation}\label{eq:payment}
I_{edge}=b_f+u_e\sum_{i=1}^{N_u}Q_i,
\end{equation}
where $b_f$ is the constant basic payment from the generation server for receiving the semantics from the edge server, $N_u$ is the number of tasks, $u_e$ is the additional fee per image quality value unit, and $Q_i$ is the value of the image quality for task $i$. The $Q_i$ is determined by different evaluation metrics according to the user's personalized requirements, which are presented in detail in Section~\ref{section:na}.

\subsubsection{Utility of the Generation Server} With consideration of the image quality $Q_i$ and (\ref{eq:payment}), the utility of the generation server can be expressed as 

\begin{equation}\label{eq:ugen1}
    U_{gen} = \sum_{i=1}^{N_u}\left(b_i+u_gQ_i\right) - I_{edge}, 
\end{equation}
where $b_i$ denotes the subscription fee paid by user $i$ for accessing this image generation service and $u_g$ represents the unit fee per quality of generated image.

\subsection{Contest in Edge Server}  
Upon receiving the payment $I_{edge}$ from the generation server, the edge server then hosts a contest use $I_{edge}$ as award. The objective of the contest in the edge server is to incentivize all semantic transfer tasks to choose a suitable transmit power to improve the final generated image quality with a fixed total payment. We define the contest as a game in which the contestants must choose a suitable transmit power to win the contest and earn an award based on semantic transfer quality. Let $P_i$ denote the power chosen by the contestant $i$. The cost function of the contestant $i$ can be defined as a twice differentiable function, i.e., $C\left(a_i,P_{i}\right)$ complying with~\cite{contest}

\begin{equation}\label{eq:1}
\left\{\begin{matrix}
    \frac{\partial C\left(a_i,P_i\right)}{\partial P_i}> 0,\\
 \frac{\partial C\left(a_i,P_i\right)}{\partial a_i}< 0,\\
 \frac{\partial^2 C\left(a_i,P_i\right)}{\partial a_i\partial P_i}< 0,
\end{matrix}\right.
\end{equation}
where $a_i$ denotes the capability of contestant $i$. The inequalities in (\ref{eq:1}) show that when the contestant has more capability (i.e., higher loss of image quality against same level semantic compression), the more likely it will consume more power to transmit the semantics. By taking (\ref{eq:1}) into consideration, we can see that the cost function can be expressed as $C\left(a_i, P_i\right)=P_i/a_i$. Sorting $N_{u}$ contestants in a descending order according to exerted effort $P_i$ to obtain the effort list $\{P_1,P_2,\ldots,P_{N_u}\}$. The contestant $i$ receives award $r_i$ for $i \in \{ 1,\ldots,N_u\}$. As the award is given in descending order, $r_1\geq r_2\geq\cdots\geq r_{N_u}$. The total award depends on the payment by the generation server and is expressed as
\begin{equation}\label{eq:r}
r_{Total}=\sum_{i=1}^{N_u}r_i=I_{edge}.
\end{equation}
The capability of each semantic transfer task is defined by the robustness of its semantics against compression loss. Referring to (\ref{capacity}), we model the capability $a_i$ as~\cite{wang2023semantic}:
\begin{equation}
    a_i=\frac{1}{Q_i\left(Z\left(D\left(P_i\right)\right)\right)}.
\end{equation}
This encourages transfer tasks with semantic that is more sensitive to compression to compete for higher transmit power to maintain quality.

We model the utility of each semantic transfer task to reflect the award from successful data transmission, as well as the costs associated with power consumption:
\begin{equation}
    U_{task}\left(a_i, P_i\right) = r_i - C\left(a_i, P_i\right).
\end{equation}
Thus, the expected utility can be derived as follows:
\begin{equation}
    \mathbb{E}\left(U_{task}\left(a_i, P_i\right)\right) = \mathcal{M}\left(a_i,r_i\right) - C_i\left(a_i, P_i\right),
\end{equation}
where $\mathcal{M}\left(a_i,r_i\right)$ denotes the expectation of the award that the contestant $i$ receives.

Accordingly, each semantic transfer task has access to its capability $a_i$ without knowing other tasks' capabilities. However, each contestant knows the distribution of the capabilities of the population, e.g., from the historical statistics~\cite{wang2023semantic}. The cumulative distribution of capability in the population of contestants is represented by a continuous function $\mathcal{P}\left(a_n\right)$. Here, we consider that $\mathcal{P} $ follows a uniform distribution. The considered distribution is easily extensible to other distributions~\cite{wang2023semantic}. For the contestant $i$, the probability that the other contestant's capability is larger than its capability is 
\begin{equation}
\label{eq:CDF:P}
{\mathcal{P}}\left({a_i}\right) = \begin{cases}
\frac{{{a_i}_{max}}-a_i}{{a_i}_{max}}, & 0\leq a_i \leq {a_i}_{max}, \\
0, & {\rm{otherwise}},
\end{cases}
\end{equation}
where ${a_i}_{max}$ is the maximum capability allowable. With the obtained $\mathcal{P}\left(a_i\right)$ and the fixed payment pool $I_{edge}$, the expected number of contestants that win before the contestant $i$ is $\left(N_T-1\right)\left(1-\mathcal{P}\left(a_i\right)\right)$. Since the payment pool is fixed, the expected payment received by the contestant $i$ is a function of the probability of winning the $i$th award (i.e., $i-1$ channels are noisier for semantic transfer tasks) multiplied by the respective $i$th award. Therefore, the expected award received by the contestant $i$ can be calculated as follows~\cite{contest}:
 \begin{equation}
 \begin{aligned}\label{eq:expect}
  \mathcal{M}\left(a_i,r_i\right)=& \sum_{i=1}^{N_{u}}r_{i}\binom{N_{u}-1}{i-1}{\mathcal{P}}^{N_{u-i}}\left(a_i\right)\left(1-\mathcal{P}\left(a_i\right)\right)^{i-1}.
\end{aligned}
\end{equation}
Then we can obtain the optimal effort in each semantic transfer task's view from
\begin{equation}\label{optimaleffort}
P_i^ *  \in \min \left( {\mathbf{P}} \right) = \min \left( {\mathop {\arg \max }\limits_{{P_i}} \;\mathcal{M}\left( {{a_i}\left( {{P_i}} \right),{r_i}} \right)} \right),
\end{equation}
where $\mathbf{P}$ is the set of optimal efforts (i.e., transmit power) that the edge server chooses for each semantic transfer task.

\subsection{Optimal Award Scheme for Contest}

The goal of the optimal award scheme is to design an award distribution that maximizes the total utility of all contestants participating in processing the semantic transfer tasks in the edge server. This requires balancing the awards based on the effort (transmit power) exerted by each contestant and their capabilities to ensure optimal system performance. The objective function for this optimization problem is defined as:
\begin{subequations}
\begin{align}
& \underset{[r_1,r_2,\ldots,r_{N_u}]}{\max} & & \sum_{i=1}^{N_u}U_{task}\left(a_i, P_i\right)=\sum_{i=1}^{N_u}\left( r_i - \frac{P_i}{a_i} \right) \label{contestopti} \\
& \quad\quad{{\rm{s.t.}}} & & \sum_{i=1}^{N_u} r_i = I_{edge}, \label{rsum} \\
& & & r_i \geq 0 \quad \forall i, \label{rnonneg} \\
& & & \sum_{m=1}^{N_u} P_{m} \leq P_{total}, \label{total_pi} \\
& & & P_{out}\left(P_i\right) \leq \theta, \quad \forall i \label{outage1} \\
& & & \left(\ref{optimaleffort}\right), \label{optimaleffortsub}
\end{align}
\end{subequations}
where (\ref{rsum}) limits the total awards not to exceed the total payment received from the generation server, (\ref{rnonneg}) ensures that each contestant receives a non-negative award, and (\ref{optimaleffortsub}) incentivizes contestants to exert optimal effort by ensuring that the award structure aligns with their power consumption and capabilities. The constraint (\ref{outage1}) ensures that the outage probability does not exceed a predetermined threshold $\theta$, and the (\ref{total_pi}) ) constraints the total transmit power for the edge server. 
The implementation of the optimal award scheme involves assessing the capability \( a_i \) of each contestant, determining the optimal effort \( P_i \) for each contestant, calculating the awards \( r_i \) based on the solution of the optimization problem and distributing the awards to the contestants accordingly. 

\subsection{Optimal Payment Design}
The goal of the contract-inspired contest framework is to achieve an optimal payment plan that maximizes total generated image quality in the generation server. This involves complex coordination and strategic planning across both servers, guided by a comprehensive set of system constraints and performance metrics. The objective function aims to maximize the total image quality by maximizing the utility of the generation server:

\begin{subequations}
\begin{flalign}
& \underset{u_e,u_g,b_i,b_f}{\max}
& &  U_{gen} = \sum_{i=1}^{N_u}\left(b_i+u_gQ_i\right) - I_{edge}\left(b_f,u_e\right), \\
& \quad\quad{{\rm{s.t.}}}
& &\sum_{i=1}^{N_u}\left(b_i+u_gQ_i\right) -  I_{edge} \geq  U_{th}\label{totalrpay}
\end{flalign}
\end{subequations}
where (\ref{totalrpay}) limits the payment to the edge server to ensure utility for the generation server above a set threshold $ U_{th}$.

\subsection{Overall Objective Function}
The core problem addressed in this framework is to optimize the overall system's performance by balancing payment design, award distribution, and resource allocation among semantic transfer tasks to achieve the highest possible image quality in a dynamic and resource-constrained environment. The overall objective function integrates the various aspects of the contract-inspired contest theoretic framework to enhance the image generation process in mobile edge immersive Metaverse applications. The objective function aims to maximize the overall quality of the generated image while ensuring efficient resource allocation.

The overall objective function can be expressed as follows:

\begin{subequations}
\begin{align}
& \underset{b_i,u_g,[r_1,r_2,\ldots,r_{N_u}]}{\max} & & \sum_{i=1}^{N_u}Q_i\left(b_i,u_g,[r_1,r_2,\ldots,r_{N_u}]\right), \label{overallutility} \\
& \quad \quad\quad{{\rm{s.t.}}} & & P_{out}\left(P_i\right) \leq \theta, \quad \forall i, \label{outage} \\
& & & \sum_{m=1}^{N_u} P_{m} \leq P_{total}, \label{totalpi} \\
& & & U_{gen} \geq U_{th}, \label{rational1} \\
& & & r_i > \frac{P_i}{a_i} \label{Uedge} \\
& & & r_i \geq 0, \forall i, \label{nonneg} \\
& & & \left(\ref{optimaleffort}\right) \label{overalloptimal}
\end{align}
\end{subequations}
where (\ref{rational1}) is the Individual Rationality (IR) constraint that keeps the generation server utility above threshold $U_{th}$, (\ref{Uedge}) guarantees that the edge server will not incur a utility loss by distributing more awards than that it receives, and (\ref{nonneg}) ensures that each contestant (i.e., a semantic transfer task) receives a non-negative award, ensuring that participating in the contest is at least as good as not participating at all. The (\ref{overalloptimal}) is the IC constraint which ensures that each semantic transfer task will choose the optimal effort (transmit power) to maximize its utility given its capability $a_i$.

The framework optimizes the overall system through three interconnected stages. First, the incentive in generation server sets the total payment to the edge server using contract theory, aligning incentives to optimize generation server utility. This total payment becomes the award pool used by the contest in edge server to incentivize edge devices to choose optimal transmit power, maximizing edge utility tied to image quality $Q_i$. Finally, the overall objective function integrates both the award distribution and payment design to maximize the overall image quality $Q_i$ under the constraints defined in the previous stages.

\section{Deep Reinforcement Learning with Generative Diffusion Model}\label{S5}

In this section, we adopt a novel method that integrates DRL with GDM to solve the optimization problem aforementioned in (\ref{overallutility}).

\subsection{Generative Diffusion Models}

\begin{figure}[t!]
\centering
\includegraphics[width=0.46\textwidth]{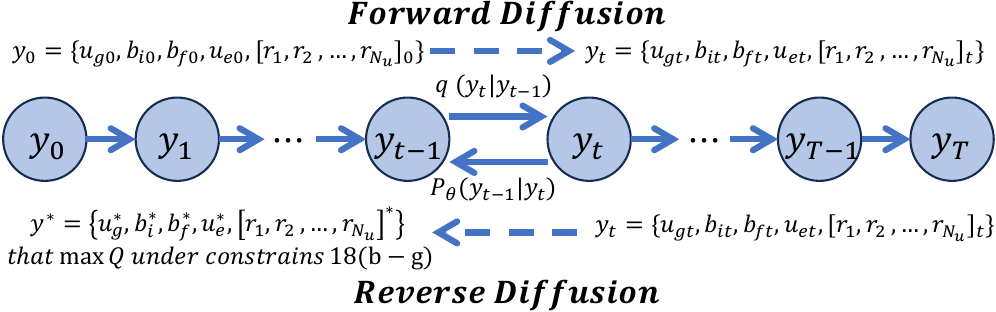}
\caption{Illustration of the forward and reverse diffusion processes: The forward diffusion process introduces gaussian noise to the current training data. In contrast, the reverse diffusion process, known as "denoising," focuses on reconstructing the original data or target data under conditions.}
\label{fig:gdms}
\end{figure}

As shown in Fig.~\ref{fig:gdms}, the basic process of GDM includes a forward diffusion phase where Gaussian noise is incrementally added to the data. This sets the stage for the reverse denoising phase, where the model learns to remove the noise and revert to the original data structure. This cycle, described as a series of probabilistic steps, highlights the ability of the model to generate and refine data through its layered approach~\cite{ho2020denoising}. The reverse diffusion process allows for the generation of new parameters of the proposed framework. In the following, we show the mechanisms of forward diffusion and reverse denoising processes, utilizing the original data $y_0$, which include the payment plan parameter and the award schemes $y_0=\{u_g,b_i,b_f,u_e,[r_1,r_2,\ldots,r_{N_u}]\}$.

1) \textbf{Forward Diffusion:} Modeled as a Markov chain with $T$ steps, the forward process incrementally adds noise, which is governed by predefined variances. Let $y_0$ denote the original payment plan, Gaussian noise with a variance of $\beta_t$  at step $t$ is added to $y_{t-1}$ to produce $y_t$ with the distribution $q\left(y_t | y_{t-1}\right)$. This process is shown as
\begin{equation}
q\left(y_t | y_{t-1}\right) = \mathcal{N}\left(y_t; \mu_t = \sqrt{1 - \beta_t} y_{t-1}, \Sigma_t = \beta_t I \right),
\end{equation}
where \(q\left(y_t | y_{t-1}\right)\) is a normal distribution defined by the mean \(\mu_t\) and variance \(\Sigma_t\), with \(I\) being the identity matrix, indicating equal standard deviation \(\beta_t\) across all dimensions.

The posterior probability from the original payment plan \(y_0\) to the final state \(y_T\) can be obtained in a manageable form as:
\begin{equation}\label{addnoise}
q\left(y_{1:T} | y_0\right) = \prod_{t=1}^T q\left(y_t | y_{t-1}\right).
\end{equation}

However, as indicated in (\ref{addnoise}), sampling \(y_t\) involves multiple calculations, which become computationally expensive for large \(t\). To mitigate this, we adopt the following updates~\cite{du2024enhancing}:

\begin{equation}\label{t}
\alpha_t = 1 - \beta_t \quad \text{and} \quad \bar{\alpha}_t = \prod_{j=0}^t \alpha_j,
\end{equation}
where \(y_t\) can then be reformulated as:

\begin{equation}
\begin{aligned}
y_t &= \sqrt{1 - \beta_t} y_{t-1} + \sqrt{\beta_t} \epsilon_{t-1} \notag \\
&= \sqrt{\alpha_t} y_{t-2} + \sqrt{1 - \alpha_t} \epsilon_{t-2} \notag \\
&= \cdots \notag \\
&= \sqrt{\bar{\alpha}_t} y_0 + \sqrt{1 - \bar{\alpha}_t} \epsilon_0,
\end{aligned}
\end{equation}
where \(\epsilon_0, \ldots, \epsilon_{t-1} \sim \mathcal{N}\left(0, I\right)\). As a result, \(y_t\) follows the distribution of
\begin{equation}
y_t \sim q\left(y_t | y_0\right) = \mathcal{N}\left(y_t; \sqrt{\bar{\alpha}_t} y_0, \left(1 - \bar{\alpha}_t\right) I\right).
\end{equation}
The forward diffusion process allows us to characterize the expected value and variance of \(y_t\). Specifically, the expected value of \(y_t\) can be expressed as:

\begin{equation}
\mathbb{E}[y_t] = \sqrt{\bar{\alpha}_t} y_0,
\end{equation}
which illustrates that \(y_t\) is influenced by the original data \(y_0\) scaled by \(\sqrt{\bar{\alpha}_t}\). $\bar{\alpha_t}$ is set to decrease as \(t\) increases, leading the expectation of \(y_t\) to gradually diminish. This indicates that the influence of the original data becomes weaker with each step of the diffusion.

Similarly, the variance of \(y_t\) is given by:

\begin{equation}
\text{Var}[y_t] = (1 - \bar{\alpha}_t) I,
\end{equation}
where the variance term \((1 - \bar{\alpha}_t)\) reflects the accumulated noise added over time, with \(I\) being the identity matrix, ensuring isotropic noise. As \(t\) progresses, \(\bar{\alpha}_t\) decreases, resulting in an increase in the variance of \(y_t\), showing that \(y_t\) is progressively dominated by Gaussian noise as the forward diffusion advances. This is a fundamental characteristic of the forward diffusion process, transforming \(y_0\) from a structured representation to a noisier state as \(t\) approaches \(T\).

2) \textbf{Reverse Diffusion Process:} This phase involves modelling the reverse transitions to retrieve the original data $y_0$ from its noised data $y_T$ by learning a series of reverse transitions. $y_T$ approximates an isotropic Gaussian distribution when $T$ is large~\cite{ho2020denoising}. The model estimates the transition probabilities using a parameterized approach, optimizing each step to reduce the divergence from the actual data distribution. We can infer the reverse distribution \(q\left(y_{t-1} | y_t\right)\) by first sampling \(y_T\) from \(\mathcal{N}\left(0, I\right)\), and then performing the reverse process to generate a sample from \(q\left(y_0\right)\). However, directly computing \(q\left(y_{t-1} | y_t\right)\) involves complex calculations with the data distribution, which is computationally impractical. To address this, we aim to approximate \(q\left(y_{t-1} | y_t\right)\) using a parameterized model \(p_\theta\), defined as follows:
\begin{equation}
p_\theta\left(y_{t-1} | y_t\right) = \mathcal{N}\left(y_{t-1}; \mu_\theta\left(y_t, t\right), \Sigma_\theta\left(y_t, t\right)\right).
\end{equation}

Following this approach, the transition from \(y_T\) to \(y_0\) can be obtained as:
\begin{equation}
p_\theta\left(y_{0:T}\right) = p_\theta\left(y_T\right) \prod_{t=1}^T p_\theta\left(y_{t-1} | y_t\right).
\end{equation}
By conditioning on timestep \(t\), the model learns to predict the Gaussian parameters, specifically the mean \(\mu_\theta\left(y_t, t\right)\) and the covariance matrix \(\Sigma_\theta\left(y_t, t\right)\) at each step. The training process for the GDMs involves optimizing the negative log-likelihood of the training data. According to \cite{ho2020denoising}, incorporating conditional information such as \(g\) into the denoising process allows \(p_\theta\left(y_{t-1} | y_t, g\right)\) to be modeled as a noise prediction function, with the covariance matrix fixed as: \begin{equation} \Sigma_\theta\left(y_t, g, t\right) = \beta_t I, \end{equation} and the mean determined by: \begin{equation} \mu_\theta\left(y_t, g, t\right) = \frac{1}{\sqrt{\alpha_t}} \left( y_t - \frac{\beta_t}{\sqrt{1 - \bar{\alpha}_t}} \epsilon_\theta\left(y_t, g, t\right)\right). \end{equation} The reverse diffusion chain, parameterized by \(\theta\), is constructed by initially sampling \(y_T \sim \mathcal{N}\left(0, I\right)\), and then performing the reverse transitions step by step: 
\begin{equation}
y_{t-1} | y_t = \frac{y_t}{\sqrt{\alpha_t}} - \frac{\beta_t}{\sqrt{\alpha_t \left(1 - \bar{\alpha}_t\right)}} \epsilon_\theta\left(y_t, g, t\right) + \sqrt{\beta_t} \epsilon,
\end{equation}
where $\epsilon \sim \mathcal{N}\left(0, I\right)$ and $t = 1, \ldots, T$. 

The original loss function for the variational lower bound can be expressed as:
\begin{equation}
L_{t} = \mathbb{E}_{y_0, t, \epsilon} \left[ w_t \| \epsilon - \epsilon_{\theta}\left(\sqrt{\bar{\alpha}_t} y_0 + \sqrt{1 - \bar{\alpha}_t} \epsilon, t, g\right) \|^2 \right],
\end{equation}
where \( w_t \) is a weighting term that depends on the variance schedule. This term is used to balance the importance of different diffusion steps. However, the study in ~\cite{ho2020denoising} proved that $w_t$ can often be disregarded without affecting model performance because it simplifies the training objective while maintaining effective noise prediction. Treating all diffusion steps uniformly by removing $w_t$ still allows the model to approximate the reverse diffusion chain accurately.

By disregarding the weighting term \( w_t \), the simplified loss function becomes~\cite{ho2020denoising}:
\begin{equation}
{L_t}' = \mathbb{E}_{y_0, t, \epsilon} \left[ \| \epsilon - \epsilon_\theta\left(\sqrt{\bar{\alpha}_t} y_0 + \sqrt{1 - \bar{\alpha}_t} \epsilon, t, g\right) \|^2 \right].
\end{equation}

The integration of GDMs and DRL is pivotal in addressing the complex demands of mobile edge immersive Metaverse image generation. The following details the formulation of states, actions, awards, and subgoals in the DRL framework and the methodological integration with GDMs.
\subsection{Integration of GDMs with DRL}

\begin{itemize}

 \item\textbf{States:} The state space in our DRL framework comprises the current resource allocations for semantic transmission task $\{P_1, P_2,\ldots, P_{N_u}\}$ and the performance metrics $\{Q_1, Q_2, \ldots, Q_{N_u}\}$ from the image generation models. In addition, environmental factors affect transmission such as the channel bandwidth $\{B_1, B_2, \ldots, B_{N_u}\}$, and the outage probability threshold $\{\theta_1, \theta_2, \ldots, \theta_{N_u}\}$ are also included.

 \item\textbf{Actions:} The action space comprises the adjustments to the parameters of the contract-inspired contest framework $\{u_g,b_i,b_f,u_e\}$, and the award allocation scheme $[r_1, r_2, \ldots, r_{N_u}]$.

\item \textbf{Rewards:} The reward structure aims to maximize the quality of generated images defined as :
\begin{equation}\label{r_g}
   R_g = \sum_{i=1}^{N_u} Q_{N_u}. 
\end{equation}
\end{itemize}

\begin{algorithm}[t!]
\caption{Optimization via Integration of GDMs with DRL}
\label{alg:training}
\begin{algorithmic}[1]
\STATE \textbf{Input:} Diffusion steps \(M\), batch size \(N_b\), discount factor \(\lambda\), soft target update parameter \(\eta\), exploration noise \(\zeta\)
\STATE \textbf{Initialize:} Replay buffer \(\mathcal{B}\), parameter network \(\epsilon_{\psi}\) with weights \(\psi\), utility network \(Q_{\chi}\) with weights \(\chi\), target parameter network \(\epsilon'_{\psi'}\) with weights \(\psi'\), target utility network \(Q'_{\chi'}\) with weights \(\chi'\)
\STATE \textbf{Initialize:} Random process \(M\) for parameter exploration

\FOR{episode = 1 to max\_episode}
    \FOR{step = 1 to max\_step}
        \STATE Observe the current environment \(e_t\)
        \STATE Set action parameters \(A_t^M\) as Gaussian noise
        \STATE Generate action parameters \(A_t^0\) by denoising \(A_t^M\) using \(\epsilon_{\psi}\)
        \STATE Add exploration noise \(\zeta\) to \(A_t^0\)
        \STATE Execute action \(A_t^0\) and observe utility \(Q_i\left(b_i,u_g,[r_1,r_2,\ldots,r_{N_u}]\right)\)
        \STATE Store the record \(\left(e_t, A_t^0, Q_i\left(b_i,u_g,[r_1,r_2,\ldots,r_{N_u}]\right)\right)\) in replay buffer \(\mathcal{B}\)
        \STATE Sample a random minibatch of \(N_b\) records \(\left(e_i, A_i,Q_i\left(b_i,u_g,[r_1,r_2,\ldots,r_{N_u}]\right)\right)\) from \(\mathcal{B}\)
        \STATE Set \(z_i = U_{gen, i} + \lambda Q'_{\chi'}\left(e_i, A_i'\right)\) where \(A_i'\) is obtained using \(\epsilon'_{\psi'}\)
        \STATE Update the utility network by minimizing the loss:
        \STATE \(L = \frac{1}{N_b} \sum_i^{N_b} \left(z_i - Q_{\chi}\left(e_i, A_i\right)\right)^2\)
        \STATE Update the parameter network by computing the policy gradient:
        \STATE \(\nabla_{\psi}\epsilon_{\psi} \approx \frac{1}{N_b} \sum_i^{N_b} \nabla_{A_i} Q_{\chi}\left(e_i, A_i\right) \nabla_{\psi} \epsilon_{\psi}\)
        \STATE Update the target networks:
        \STATE \(\psi' \leftarrow \eta \psi + \left(1 - \eta\right) \psi'\)
        \STATE \(\chi' \leftarrow \eta \chi + \left(1 - \eta\right) \chi'\)
    \ENDFOR
\ENDFOR
\STATE \textbf{Return:} The trained parameter network \(\epsilon_{\psi}\)

\end{algorithmic}
\end{algorithm}

The integration of GDMs with DRL in our framework serves to enhance both the quality and efficiency of resource allocation in mobile edge immersive Metaverse environments. This approach leverages the strengths of GDMs in generating high-quality parameter samples and the adaptability of DRL in optimizing decisions within dynamic environments.

\textbf{Policy Refinement via DRL:} As shown in Algorithm~\ref{alg:training}, the training process of the integrated GDMs and DRL differs significantly from traditional DRL models including PPO and SAC. Traditional models optimize policy parameters through gradient updates directly, whereas our framework leverages GDMs to iteratively refine both the action space and reward structures. This approach enhances exploration, allowing the model to maintain broader exploration capacity while achieving faster convergence. Within the DRL framework, the reward function~\ref{r_g} plays a significant role in guiding the agent towards actions that maximize the generated image quality and resource efficiency.

In each time step, the reward value is updated based on the current environment and the actions taken. This iterative reward update mechanism is crucial for the agent to learn from the environment, enhancing the decision-making process over time. The reward update equation is defined as:

\begin{equation}
    R_{g_{t+1}} = R_{g_{t}} + \kappa\nabla Q_{\chi}(e_t, A_t).
\end{equation}

Here, $R_{g_{t}}$ represents the reward at time step $t$, and the term $\kappa \nabla Q_{\chi}(e_t, A_t)$ represents the learning adjustment, where $\alpha$ is the learning rate and $\nabla Q_{\chi}$ is the gradient of the reward estimation network. This iterative process helps the agent learn from the environment, enhancing decision-making over time. This description emphasizes how GDM integrated DRL differs from PPO and SAC in exploration and refinement, making it better suited for complex, dynamic environments.

\textbf{Parameter Generation via GDMs:} The proposed approach leverages GDMs to generate high-quality parameter samples that serve as inputs for the DRL framework. The parameter network \(\epsilon_{\psi}\) is trained to generate appropriate action parameters that maximize the utility function while considering the constraints of the mobile edge environment.

To train the parameter network, the back-propagation algorithm is employed to adjust the network weights based on the error between the predicted and target utility values. The gradient of the loss function with respect to the model parameters is given by:

\begin{equation}
    \nabla_{\psi} L = \frac{1}{N_b} \sum_i^{N_b} \left( Q_{\chi}(e_i, A_i) - z_i \right) \nabla_{\psi} f_{\psi}(e_i).
\end{equation}

Here, $\nabla_{\psi} L$ denotes the gradient of the loss function $L$, $Q_{\chi}(e_i, A_i)$ is the utility estimation from the network, and $z_i$ represents the target value obtained through experience replay. The replay buffer size is denoted by $N_b$. This equation highlights how the error in predicted utility guides parameter updates, thereby reducing the overall loss and ensuring convergence towards the optimal policy.

The back-propagation process, along with parameter generation via GDMs, ensures that the DRL agent can effectively learn and adapt to changes in network conditions. The generated parameters serve as a starting point, and the DRL framework iteratively refines these suggestions based on environmental feedback, achieving faster convergence and greater stability compared to traditional methods.

\textbf{Inference Sampling:} The inference process in the proposed framework is distinct from more traditional approaches like PPO and SAC, where deterministic or stochastic policies are used for action selection. Instead, the proposed framework utilizes the diffusion model to generate the optimal parameters through a denoising process. Algorithm~\ref{alg:inference} presents the pseudo-code for inference sampling, following a reverse diffusion process to reconstruct the original parameter settings from Gaussian noise. The generated parameters are subsequently used to guide the action selection in the DRL component, effectively optimizing the resource allocation for image generation.

\begin{algorithm}[t!]
\caption{Inference Sampling via Reverse Diffusion}
\label{alg:inference}
\begin{algorithmic}[1]
\STATE \textbf{Input:} Number of diffusion steps \(T\), trained parameter network \(\epsilon_{\psi}\)
\STATE \textbf{Initialize:} \(A_T \sim \mathcal{N}(0, I)\) \COMMENT{Sample from standard normal distribution}
\FOR{\(t = T, T-1, \ldots, 1\)}
    \STATE Generate noise \(z \sim \mathcal{N}(0, I)\) if \(t > 1\), else set \(z = 0\)
    \STATE Compute \(A_{t-1}\) using:
    \begin{equation*}
   A_{t-1} = \frac{A_t}{\sqrt{\alpha_t}} - \frac{\beta_t}{\sqrt{\alpha_t (1 - \bar{\alpha}_t)}} \epsilon_\psi(A_t, t) + \sqrt{\beta_t} z,
    \end{equation*}
\ENDFOR
\STATE \textbf{Return:} \(A_0\) \COMMENT{${u_g}^*,{b_i}^*,{b_f}^*,{u_e}^*, [r_1, r_2, \ldots, r_{N_u}]^*$}
\end{algorithmic}
\end{algorithm}

The inference sampling procedure utilizes the trained parameter network \(\epsilon_\psi\) to progressively denoise an initial Gaussian noise sample \(A_T\). Beginning with \(A_T\), which is sampled from a standard normal distribution, the reverse diffusion process is used to reconstruct the original parameter settings \(A_0\) by gradually removing noise. The trained network \(\epsilon_\psi(A_t, t)\) predicts the noise at each step, enabling us to reverse the forward diffusion and recover the original parameter state. Specifically, the reverse diffusion step is given by:

\begin{equation}
A_{t-1} = \frac{A_t}{\sqrt{\alpha_t}} - \frac{\beta_t}{\sqrt{\alpha_t (1 - \bar{\alpha}_t)}} \epsilon_\psi(A_t, t) + \sqrt{\beta_t} z,
\end{equation}
where \(z \sim \mathcal{N}(0, I)\) for \(t > 1\), and \(z = 0\) for \(t = 1\). This iterative denoising process ensures that we obtain the optimal parameter configuration \(A_0\), which can be subsequently used in the DRL optimization. This method effectively refines the parameters, ensuring their suitability for subsequent decision-making tasks.

\textbf{Dual Optimization Process:} The integration framework operates through a dual optimization process, where the contract-inspired contest theory drives the incentive design, and DRL fine-tunes this design based on observed outcomes, leveraging GDMs for iterative refinement.

\begin{itemize}
\item \textit{Contest Scheme Optimization:} The contract-inspired contest theory provides an initial design for the reward distribution $[r_1, r_2, \ldots, r_{N_u}]$ and the total incentive pool $I_{edge}$, using GDMs to iteratively refine these parameters. The diffusion process allows for the exploration of different possible reward structures, while the reverse diffusion process fine-tunes the reward allocations to ensure optimal utility for each task. The reward distribution in this context is dynamic—it is iteratively updated to reflect the changing environment, as influenced by DRL adjustments based on the observed effort levels and quality of generated semantic data. Essentially, GDMs explore the space of potential reward schemes, and DRL optimizes these schemes in a feedback-driven manner to maintain incentive compatibility and maximize participation.

\item \textit{Payment Design Optimization:} While optimizing the contest scheme, the DRL framework also adjusts the overall payment parameters $\{u_g,b_i,b_f,u_e\}$, including factors such as subscription fees and additional incentives. This fine-tuning by DRL ensures that the reward is balanced among edge devices to maintain efficient resource utilization and encourage optimal participation in semantic transfer tasks. The DRL agent learns from the system's real-time feedback to adapt the payment structure to different environmental conditions, avoiding collusion while maximizing the overall quality of generated images. 
\end{itemize}

The integrated approach allows dynamic adjustments to award structures and payment designs, enhancing system robustness. However, this integration of GDMs and DRL introduces computational complexity. The GDM component has a complexity of \(O\left(Mn\right)\)~\cite{du2024enhancing}, where \(n\) represents the number of parameters or data dimensions, while the DRL component has a complexity of \(O\left(MN_b\mathcal{C}\mathcal{S}\right)\)~\cite{schulman2017proximal}. Although computationally intensive, this combined framework significantly enhances convergence speed and scalability. Here, $\mathcal{S}$ represents the number of possible states (e.g., resource allocations, environmental factors) and $\mathcal{C}$ represents the number of possible actions (e.g., parameter adjustments) within the DRL framework.

\begin{figure*}[t!]
\centering{\includegraphics[width=0.95\textwidth]{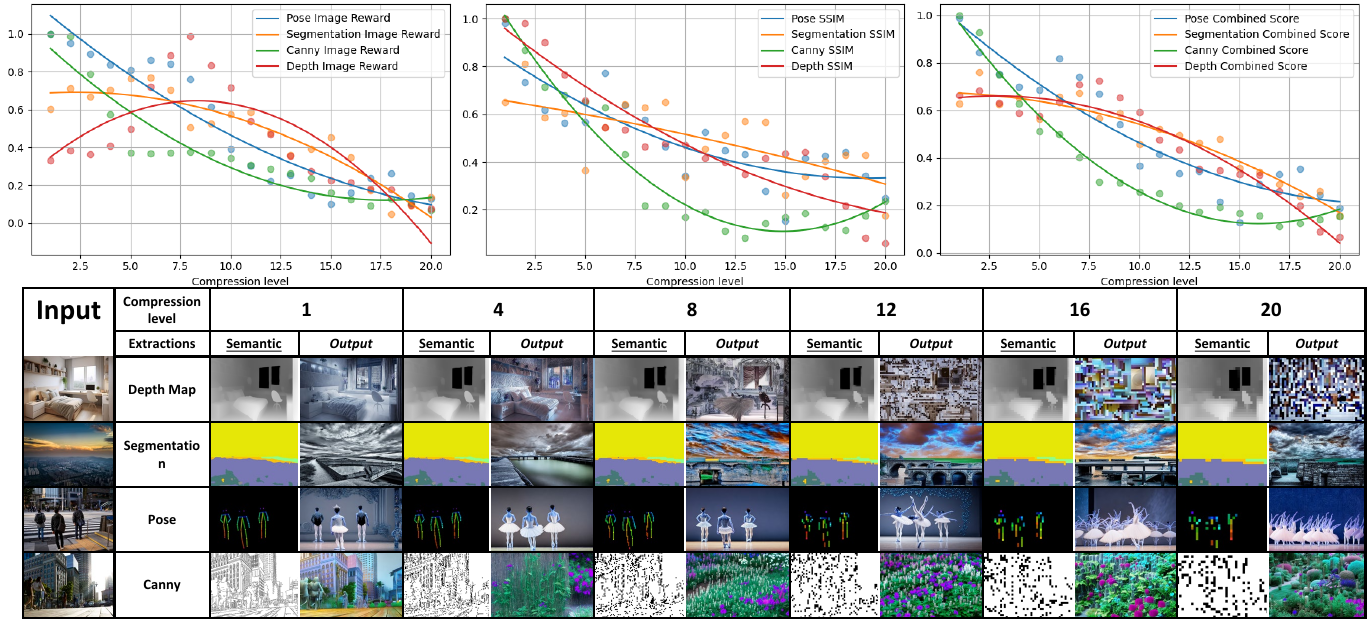}}
\caption{Image quality as a function of different levels of semantic compression for various types of semantic inputs (depth map, segmentation, pose estimation, and Canny edge detection). The results indicate a general decrease in image quality with increased compression, with depth maps showing an anomaly at certain levels.}
\label{fig:Approximation}
\end{figure*}

\section{Performance Evaluation}\label{section:na}
This section presents a comprehensive evaluation of the proposed framework by focusing on three main areas of investigation:
\begin{itemize}
    \item \textbf{Modeling the Impact of Semantic Downscaling on Generated Image Quality: } We assess how different levels of semantic compression affect the quality of the generated images in the mobile edge immersive Metaverse environment. The experiments explore various semantics including depth maps, segmentation, pose estimation, and Canny edge detection, and analyze how compression impacts image quality metrics.
    \item \textbf{Incentive Mechanism for Optimizing Transmit Power Levels: }We evaluate the effectiveness of the proposed incentive mechanism in influencing the transmit power levels of semantic transfer tasks by exploring different reward settings and analyzing how these incentives impact task behavior, particularly regarding transmit power allocation.
    \item \textbf{Comparative Analysis of Algorithm Efficiency and Effectiveness: }We compare the performance of the proposed algorithm with traditional DRL methods, focusing on key metrics such as convergence speed, stability, and overall performance. The comparison highlights the advantages of the proposed algorithm in solving the optimization problem associated with the incentive mechanism, particularly in complex environments.
\end{itemize}

\subsection{Dataset, Pretrained Model, Experiment Platform and Parameter Settings}

Our study utilizes four pretrained models from ControlNet, which are trained on the following datasets:
\begin{itemize}
    \item \textbf{LAION-5B~\cite{schuhmann2022laion}}: A large-scale dataset with 585 billion CLIP-filtered image-text pairs, crucial for robust model training.
    \item \textbf{ADE20K~\cite{zhou2017scene}}: Used for semantic segmentation tasks, providing detailed image annotations.
    \item \textbf{Light-weight OpenPose~\cite{osokin2018real}}: For human keypoint detection and pose semantic extraction.
    \item \textbf{Image-to-Image Translation Datasets}: Including Canny edges~\cite{canny1986computational}, depth maps~\cite{ranftl2020towards} for semantic extractions.
\end{itemize}
We next detail the specifications of the equipment used in our experiments. We use the following system as a testbed to simulate the edge server and generation server environments:

\begin{itemize}
\item \textbf{CPU (Edge Server)}: Intel(R) Core(TM) i7-10510U CPU @ 1.80GHz
\item \textbf{GPU (Edge Server)}: NVIDIA GeForce GTX 1650 with Max-Q Design
\item \textbf{Operating System (Edge Server)}: Windows 11
\item \textbf{CPU (Generation Server)}: AMD Ryzen Threadripper PRO 3975WX 32-Cores
\item \textbf{GPU (Generation Server)}: NVIDIA RTX A5000
\item \textbf{Operating System (Generation Server)}: Linux Ubuntu 20.04.6
\end{itemize}

The experimental topology consists of four edge devices connected to an edge server, which is then connected to a central generation server through a wireless channel. The edge server, running on Windows 11 with an Intel Core i7-10510U CPU and NVIDIA GeForce GTX 1650 GPU, performs semantic extraction from user input data. These semantics are then compressed and transmitted to the generation server, which runs on Linux Ubuntu 20.04 with an AMD Ryzen Threadripper PRO 3975WX CPU and NVIDIA RTX A5000 GPU, to generate high-quality images. 

This setup emulates a mobile edge network scenario for immersive Metaverse applications, focusing on resource constraints and network dynamics. Wireless communication parameters used to evaluate the system performance are given in Table \ref{tab:wireless_params}.

\begin{table}[H]

\centering
\caption{Wireless Communication Parameters}
\vspace{-0.2cm}
\begin{tabular}{m{4.5cm}|m{2cm}}
\toprule
\hline
\textbf{Parameter} & \textbf{Value} \\ 
\hline
Number of service users $N_u$ & $4$ \\ \hline
Bandwidth per channel $B$ & $5 MHz$ \\ \hline
Noise power $\sigma_{N}^{2}$ & $9\times10^{-6}mW$ \\ \hline
Outage probability threshold $\theta$ & $0.05$ \\ \hline
Outage probability  $\theta$ & $0.003$ \\ \hline
Transmit power range $P_i$ & $5 - 100 mW$ \\ \hline
Frame rate requirement $x$ & $30 fps$ \\ \hline
Total transmit power $P_{total}$ & $100mW$ \\ \hline
Channel gain range $|g|^2$ & $10^{-6} - 10^{-3}$  \\ \hline
\bottomrule
\end{tabular}
\label{tab:wireless_params}
\end{table}

\subsection{Image Quality Response to Semantic Compression Level}
\begin{figure*}[t!]
  \centering
    \includegraphics[width=0.95\textwidth]{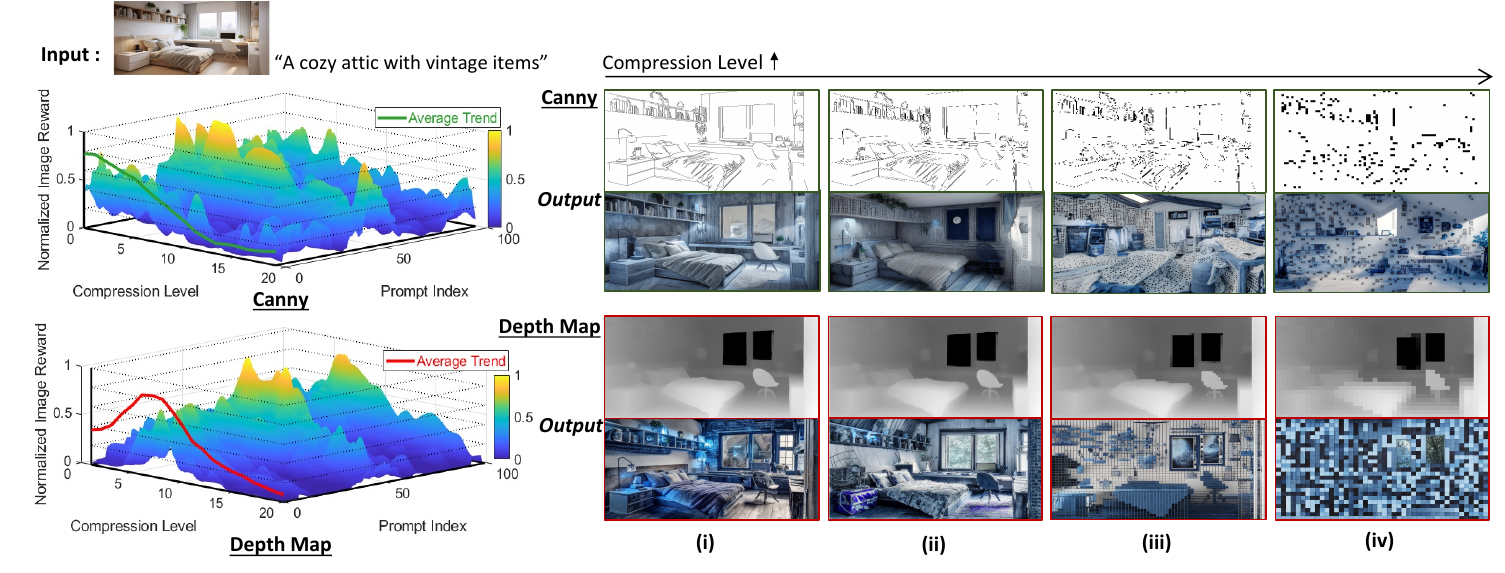}
  \caption{Detailed analysis of image quality for each prompt using depth map semantics. The anomaly in the depth map results, where a slight compression led to higher image quality(from (i) to (ii)), can be attributed to the elimination of excess details, which simplified the generated images and enhanced their overall coherence.}
  \label{fig:matlab}
\end{figure*}
To assess how different levels of semantic compression affect the quality of the generated images in the mobile edge immersive Metaverse environment, we conduct a series of experiments. Each semantic input (i.e., depth map, segmentation, pose estimation, and Canny edge detection) is compressed at different levels, ranging from 1 to 20. The experiments are repeated 100 times with randomly generated prompts, and the results were averaged to determine the relationship between the compression level and generated image quality. The metrics used to assess image quality are SSIM~\cite{brunet2011mathematical} and Image Reward~\cite{xu2024imagereward}, both normalized to a range of $[0, 1]$. Refer to~(\ref{eq:quality}), \(\beta\) is set to 0.5 to balance the semantics alignment and the text alignment but can be adjusted based on the importance of semantic alignment. 

Figure~\ref{fig:Approximation} illustrates the overall trend observed across different semantics. As the compression level increases, both SSIM and Image Reward generally decrease, indicating a loss in image quality due to the reduction in semantic information. A notable observation is that the depth map semantics exhibit an increase in image quality at certain compression levels, specifically from level 5 to 10. This result is then analyzed further as presented in Fig.~\ref{fig:matlab}.

As discussed in~\cite{yu2024promptfix}, excessive prompts can degrade the quality of the generated image by introducing unnecessary complexity. Compression helps to eliminate these superfluous details, leading to cleaner and more aesthetically pleasing output. For instance, a prompt such as ``A cozy attic with vintage items" resulted in higher image quality when the depth map semantics were slightly compressed, reducing the granularity of the generated objects. For example, lamps are generated as vases or other vintage items in Fig.~\ref{fig:matlab} Depth Map (ii), enhancing the overall scene’s cohesiveness and immersiveness Although high-fidelity semantics are generally preferred, slight compression can improve image quality by removing extraneous details that may not contribute to the desired output. 

\subsection{Incentive Mechanism Behavior Analysis}
To assess the effectiveness of the incentive mechanism in influencing transmit power allocation for different semantic transfer tasks, we conducted a series of experiments focusing on reward setting, channel gain, and total image quality. The findings are categorized below:

\subsubsection{Reward Setting Effect}
\begin{figure}[t!]
\centering{\includegraphics[width=0.46\textwidth]{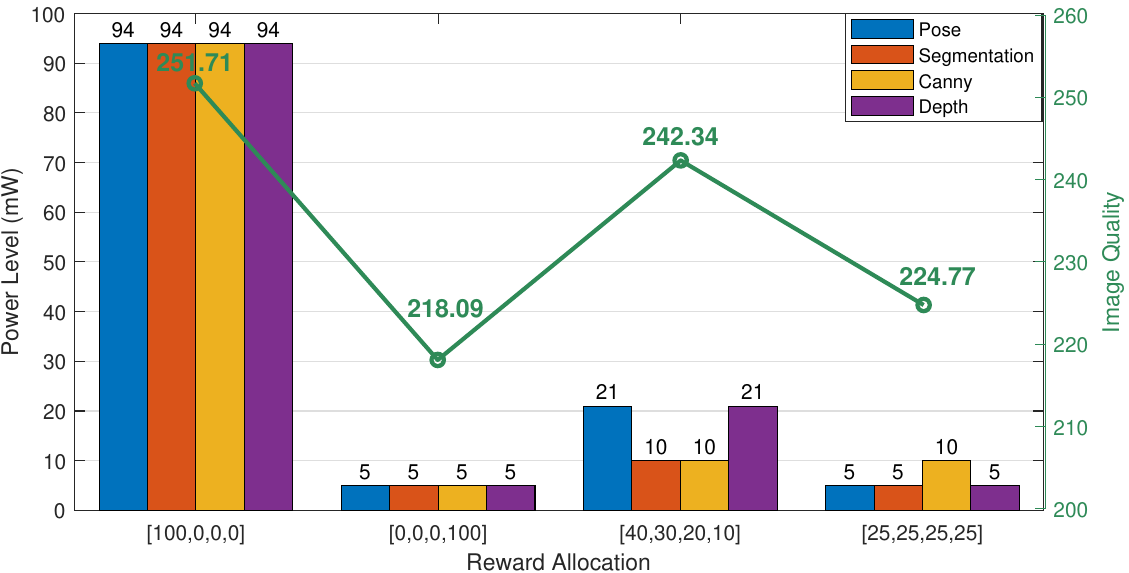}}
\caption{This figure illustrates how different reward distributions affect the transmit power levels chosen by four semantic transfer tasks (Pose, Segmentation, Canny, and Depth). The reward settings range from a winner-takes-all distribution [100,0,0,0] to an evenly distributed reward [25,25,25,25]. The results show that the allocation of rewards significantly influences the power levels and the generated image quality, with tasks adjusting their efforts based on the reward structure.}
\label{fig:reward}
\end{figure}

In this experiment, we kept the channel gain fixed at $10^{-4}$ to eliminate the randomness. The reward setting significantly impacts the choice of transmit power by different semantic transfer tasks. As shown in Fig.~\ref{fig:reward}, varying the reward distribution has notable effects:
\begin{itemize}
    \item \textbf{Winner-Takes-All Setting $[100,0,0,0]$:} When the entire reward is allocated to the first ``winner," all tasks exert maximum power (94 mW) to secure the reward, reflecting the competitive nature of this setup. This strategy effectively balances the transmission cost by incentivizing high effort.
    \item \textbf{Last-Place Reward $[0,0,0,100]$:} Conversely, when the reward is allocated solely to the last contestant, all tasks reduce their power to the maximum power (constrained by the outage probability), as the incentive is to get the last award instead of first one.
    \item \textbf{Even Distribution $[25,25,25,25]$:} When the reward is evenly distributed, each semantic transmission task tends to exert minimal effort, as there is no differential gain in winning. This results in relatively low power allocations across the board.
    \item \textbf{Gradually Decreasing Distribution $[40,30,20,10]$:} A decreasing reward distribution encourages tasks to allocate their transmit power less extreme, balancing between the reward and the effort. This setting promotes efficient resource usage while maintaining a competitive edge among the tasks.
\end{itemize}

\subsubsection{Channel Gain Effect}
\begin{figure}[t!]
\centering{\includegraphics[width=0.46\textwidth]{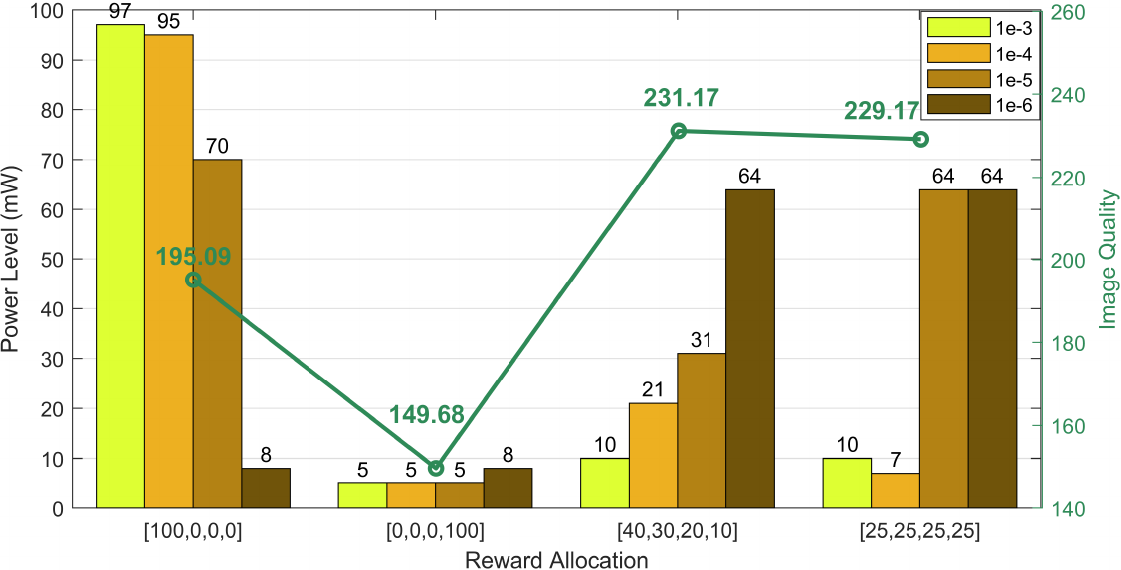}}
\caption{This figure demonstrates the effect of varying channel gains (from $10^{-3}$ to $10^{-6}$) on the transmit power levels for the same four semantic transfer tasks. The results show that lower channel gains lead to reduced power levels, particularly in settings where rewards are unevenly distributed. Tasks with lower channel gains adjust their efforts accordingly, showing the adaptability of the contest-based framework under different channel conditions.}
\label{fig:channel}
\end{figure}
To evaluate the contest theory-based framework under different channel conditions, we conducted experiments with varying channel gains while all tasks transmitted Canny edges. As illustrated in Fig.~\ref{fig:channel}, the results highlight the impact of channel gain on power allocation. Tasks with lower channel gains maintain low power levels, as increasing power does not substantially improve image quality due to poor channel conditions. This is particularly evident when channel gains are reduced to $10^{-5}$ and $10^{-6}$, where tasks remain at minimal power. Interestingly, when rewards are evenly distributed, tasks with lower channel gains tend to increase their power levels. This occurs because the contest incentivizes them to improve their performance, even under challenging conditions, to avoid being overtaken by other tasks.

\subsubsection{Contract Payment Effect}
\begin{figure}[t!]
\centering{\includegraphics[width=0.46\textwidth]{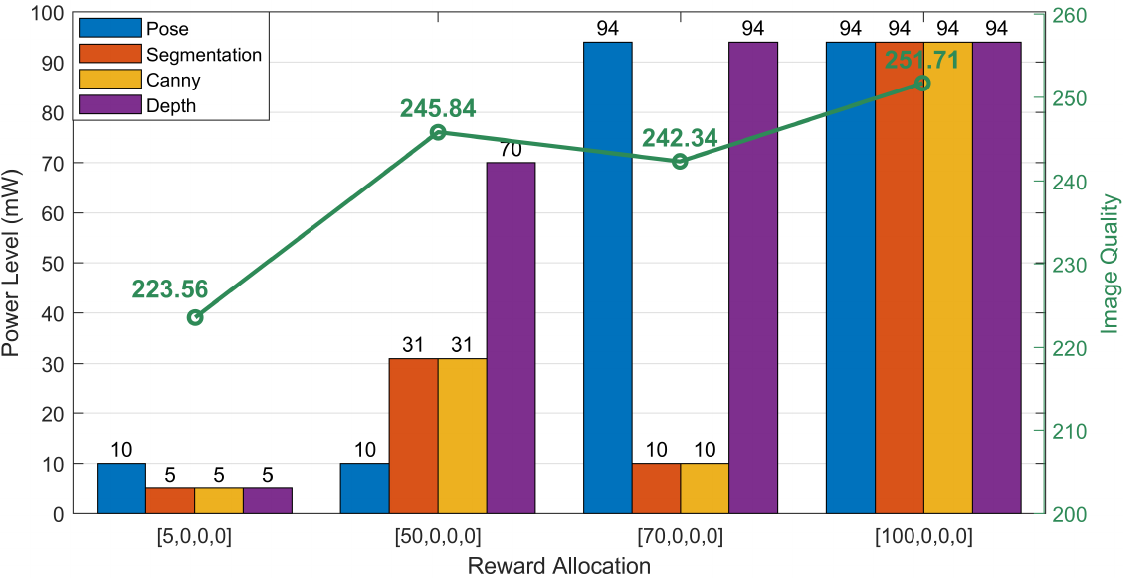}}
\caption{This figure presents the effect of different total reward levels (ranging from 5 to 100) on the transmit power allocation for the four semantic transfer tasks. As the total reward increases, tasks with higher capabilities (such as Pose and Depth) increase their power levels to maximize their share of the reward, while lower rewards result in minimal power allocation and lowest image quality across all tasks.}
\label{fig:ivalue}
\end{figure}
A common challenge in contest theory is collusion when the total reward is not fixed. To mitigate this, we introduced a contract theory-inspired mechanism that adjusts the total payment based on the contract. As shown in Fig.~\ref{fig:ivalue}, different total reward values (ranging from 5 to 100) influence the transmit power chosen by each task:
\begin{itemize}
    \item \textbf{Small Payment:} With a small payment, all tasks choose minimal power settings (5 mW), indicating that the incentive is insufficient to justify higher power consumption.
    \item \textbf{Large Payment:} As the total payment increases, tasks, particularly those with higher capabilities, increase their power levels to maximize their share of the reward. This results in higher power allocation strategies, especially in tasks such as Pose and Depth.
\end{itemize}
This contract-based adjustment ensures that tasks are incentivized to perform optimally without colluding, maintaining competitive fairness across the system.

\subsection{Performance Evaluation of the integration of GDM and DRL}
\begin{figure}[t!]
\centering
\subfloat[]{%
    \includegraphics[width=0.379\textwidth]{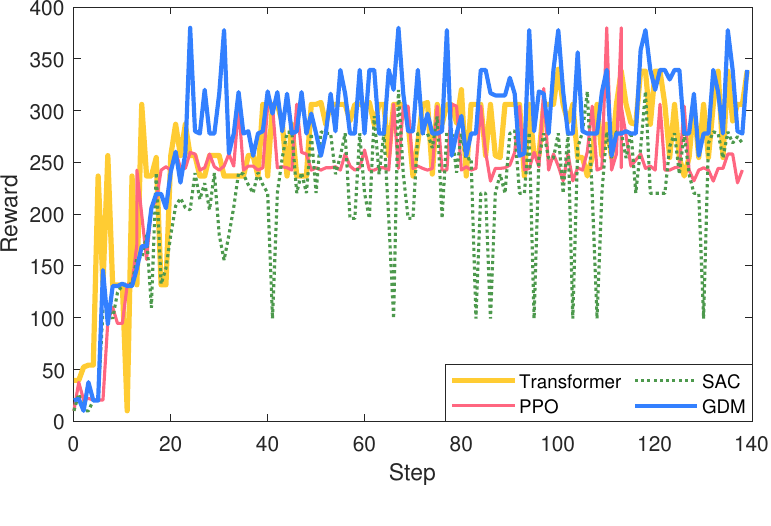}%
    \label{fig:benchmark_a}
}
\subfloat[]{%
    \includegraphics[width=0.11\textwidth]{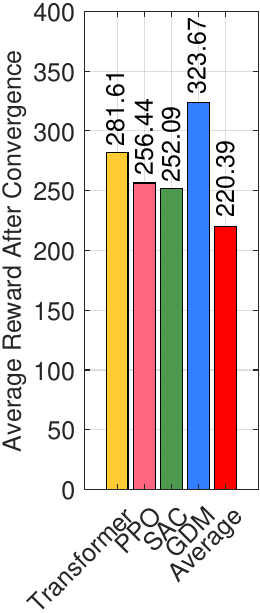}%
    \label{fig:benchmark_b}
}
\caption{\textbf{(a)} illustrate performance in terms of reward over steps. GDM demonstrates superior convergence speed and stability, achieving higher rewards consistently across steps, especially compared to PPO and SAC, which exhibit more fluctuations due to the increased complexity of image restoration with downscaling. \textbf{(b)} shows the average reward level after each method converges.}
\label{fig:benchmark}
\end{figure}

Finally, we evaluate the performance of the proposed algorithm against several traditional DRL methods. Figure~\ref{fig:benchmark} presents the benchmarking results for five different algorithms: GDM, PPO, SAC, Transformer-based SAC, and average power allocation. The PPO and SAC models are based on classic policy gradient and actor-critic frameworks, respectively, and represent commonly used techniques in DRL. Additionally, we included a transformer-based DRL model that leverages self-attention to capture dependencies across action sequences. The comparison focuses on convergence speed, stability, and overall performance.

\begin{itemize}
    \item \textbf{Convergence Speed:} The GDM algorithm demonstrates rapid convergence, reaching an average reward of 303.67 at step 25. This convergence rate matches that of PPO, which also achieves convergence at step 25 with a lower average reward of 256.94. Furthermore, GDM outperforms SAC, which converges at step 42 with an average reward of 252.09, and Transformer-based SAC, which converges at step 30 with an average reward of 281.61.
    \item \textbf{Stability:} Although SAC showed a slightly higher average reward than that of the PPO, its performance is less stable, with noticeable fluctuations. The substantial fluctuation observed within this process is a consequence of the increased complexity of image restoration when the downscaling factor increases. Specifically, an increased downscaling factor signifies a greater division of the image's width and height. This substantial partitioning invariably complicates restoring the original semantic information, thereby influencing the quality of the final generated image. In contrast, GDM provided more consistent results, making it a preferable choice for scenarios where stability is crucial.
    \item \textbf{Performance:} GDM not only demonstrates higher stability but also higher generated image quality. Upon convergence, GDM achieved an average reward of 303.67, surpassing average power allocation by $46.95\%$, PPO by $26.29\%$, SAC by $28.39\%$, and Transformer-based SAC by $14.93\%$. These results underscore GDM's efficacy in optimizing the contract-inspired contest framework, offering high-quality image generation coupled with resilient performance across varying conditions.
\end{itemize}
These results demonstrate that GDM, integrated with DRL, provides superior performance in optimizing the proposed framework, ensuring efficient resource allocation and high-quality image generation.

\section{Conclusion and Future Directions}\label{conclu}
In this paper, we have proposed a novel framework integrating contract-inspired contest theory, GDM and DRL for controllable image generation in mobile edge Metaverse environments. The proposed framework optimizes resource allocation and incentivizes edge devices to transmit high-quality semantic data, crucial for generating realistic and immersive images. By leveraging the flexibility of GDMs and the adaptability of DRL, our approach effectively addresses the dynamic challenges of resource-constrained edge networks. The experimental results demonstrated that the framework significantly improves image quality, convergence speed, and stability over traditional methods, making it particularly suitable for complex resource allocation tasks in mobile edge Metaverse applications. In future work, we plan to explore incentive mechanisms for multimodal semantic data transmission to improve image generation quality in diverse Metaverse applications. Additionally, refining reward structures and semantic extraction techniques can further enhance system performance in resource-constrained environments.
\bibliographystyle{IEEEtran}
\bibliography{Ref}

\end{document}